\documentclass[pdftex,twocolumn,epjc3]{svjour3}          

\RequirePackage[T1]{fontenc}

\smartqed  

\RequirePackage{graphicx}
\RequirePackage{mathptmx}      
\RequirePackage{flushend}
\RequirePackage[numbers,sort&compress]{natbib}
\RequirePackage[colorlinks,citecolor=blue,urlcolor=blue,linkcolor=blue]{hyperref}

\usepackage{amsmath}
\usepackage{amssymb}
\usepackage{subfigure}
\usepackage{tabularx}
\usepackage{dcolumn}
\usepackage{bm}
\usepackage{xcolor}

\journalname{Eur. Phys. J. C}

\begin{document}

\title{Isolating the confining color field in the SU(3) flux tube}

\author{M. Baker\thanksref{e1,addr1}
\and
P. Cea\thanksref{e2,addr2}
\and
V. Chelnokov\thanksref{e3,addr3,addr4}
\and
L. Cosmai\thanksref{e4,addr2}
\and
F. Cuteri\thanksref{e5,addr5}
\and
A. Papa\thanksref{e6,addr3,addr6}
}

\institute{Department of Physics, University of Washington, WA 98105 Seattle, USA\label{addr1}
\and
INFN - Sezione di Bari, I-70126 Bari, Italy\label{addr2}
\and
INFN - Gruppo collegato di Cosenza, I-87036 Arcavacata di Rende, Cosenza, Italy\label{addr3}
\and
\emph{on leave of absence from} Bogolyubov Institute for Theoretical Physics of the National Academy of Sciences of Ukraine\label{addr4}
\and
Institut f\"ur Theoretische Physik, Goethe Universit\"at, 60438 Frankfurt am Main, Germany\label{addr5}
\and
Dipartimento di Fisica dell'Universit\`a della Calabria, I-87036 Arcavacata di Rende, Cosenza, Italy\label{addr6}
}

\thankstext{e1}{e-mail: mbaker4@uw.edu}
\thankstext{e2}{e-mail: paolo.cea@ba.infn.it}
\thankstext{e3}{e-mail: volodymyr.chelnokov@lnf.infn.it}
\thankstext{e4}{e-mail: leonardo.cosmai@ba.infn.it}
\thankstext{e5}{e-mail: cuteri@th.physik.uni-frankfurt.de}
\thankstext{e6}{e-mail: alessandro.papa@fis.unical.it}

\date{Received: date / Accepted: date}

\maketitle

\begin{abstract}
Using lattice Monte Carlo simulations of  SU(3) pure
gauge theory,  we  determine the spatial distribution of all components of
the color fields created by a static quark and antiquark. We identify the
components of the measured chromoelectric field transverse to the line
connecting the quark-antiquark pair with the transverse components of an
effective Coulomb-like field $\vec{E}^C $ associated with the quark sources.
Subtracting $\vec{E}^C$ from the total simulated chromoelectric field
$\vec{E}$ yields a non-perturbative, primarily longitudinal chromoelectric
field $\vec{E}^{NP}$,  which we identify as the confining field. 
This is the first time that  the chromoelectric field  has been separated into perturbative 
and nonperturbative components, creating a new tool to
study the color field distribution between a quark and an antiquark, 
and thus the long distance force between them. \end{abstract}

\section{Introduction}
Quantum chromodynamics (QCD), the theory of the strong interactions
describing the dynamics of quarks and gluons, has yet to provide a theoretical
explanation of the experimentally established phenomenon of confinement,
{\it i.e.}, the confinement of quarks and gluons inside hadrons.
Several mechanisms of confinement have been proposed (for a review, see
Refs.~\cite{greensite2011introduction,Diakonov:2009jq}), each with its own
merits and limitations, but a comprehensive picture is still missing.
In particular, it is not yet clear which feature of QCD is responsible for
the area-law behavior of Wilson loops that implies a linear confining potential
between a static quark and antiquark at large distances. Results from numerical simulations have shown this linear potential for $q \bar q$  distances
$\gtrsim$ 0.5 fm, and up to distances of about 1.4 fm in presence of dynamical
quarks, where {\em string breaking} should take
place~\cite{Philipsen:1998de,Kratochvila:2002vm,Bali:2005fu}.

A wealth of numerical analyses of SU(2) and SU(3) Yang-Mills
theory~\cite{Fukugita:1983du,Kiskis:1984ru,Flower:1985gs,Wosiek:1987kx,DiGiacomo:1989yp,DiGiacomo:1990hc,Cea:1992sd,Matsubara:1993nq,Cea:1994ed,Cea:1995zt,Bali:1994de,Skala:1996ar,Haymaker:2005py,D'Alessandro:2006ug,Cardaci:2010tb,Cea:2012qw,Cea:2013oba,Cea:2014uja,Cea:2014hma,Cardoso:2013lla,Caselle:2014eka,Cea:2017ocq,Shuryak:2018ytg}
have found that the dominant color
field generated by a static quark-antiquark pair is the component of the
chromoelectric field along the line connecting the pair.
(See, in particular, the SU(2) studies of Ref.~\cite{Cea:1995zt}.) This
longitudinal field results in tube-like structures (flux tubes) that naturally
give rise to a long-distance linear quark-antiquark potential
~\cite{Bander:1980mu,Greensite:2003bk,Ripka:2005cr,Simonov:2018cbk}.

The aim of this paper is to measure the complete color field distributions generating this
heavy quark potential. To do this, we first
perform a series of new simulations in SU(3) pure gauge theory, measuring all
six components of the color electric and magnetic fields on all transverse planes passing
through the line between the quarks.
These simulations have been carried out for three values of the quark-antiquark separation, and they provide maps of the chromodynamic fields permeating the space between a quark and antiquark.  (These fields can be viewed as analogous to the electromagnetic field permeating the space between a pair of oppositely charged particle obtained by the solution of Maxwell's equations.)

We find that the chromomagnetic field is everywhere
much smaller than the chromoelectric field. We then fit the measured
{\em transverse} components of the chromoelectric field to an effective
Coulomb-like field generated by sources at the positions of the quarks.
A {\em nonperturbative}, mostly longitudinal, chromoelectric field is then
obtained by subtracting the effective Coulomb-like field from the total
chromoelectric field, thereby isolating its confining part.
To the extent that the nonperturbative field generates the  measured linear term in the long distance heavy quark potential,  
and the effective Coulomb-like field generates the measured Coulomb correction to the force, we will have gained  
new understanding into the development of the long distance force between a quark and an antiquark
in terms  of the color fields permeating the space between them.

\section{Theoretical background and lattice observables}

\begin{figure}[t]
   \centering
   \subfigure[\hspace{2cm}]%
        {\label{fig:operator_Wilson}\includegraphics[width=0.4\textwidth,clip]{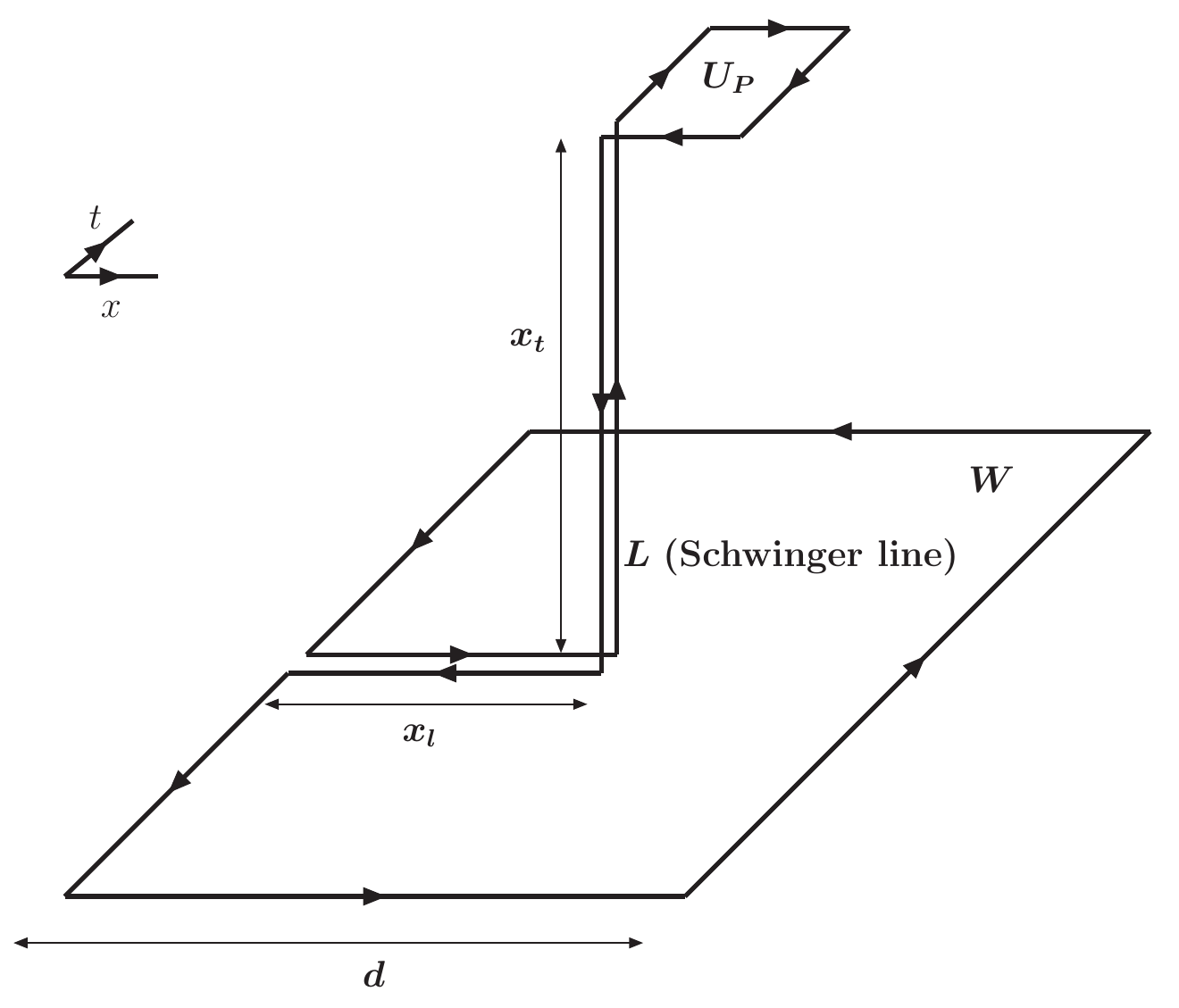}}\hspace{-2cm}
   \subfigure[]%
        {\label{fig:qqbar}\includegraphics[width=0.15\textwidth,clip]{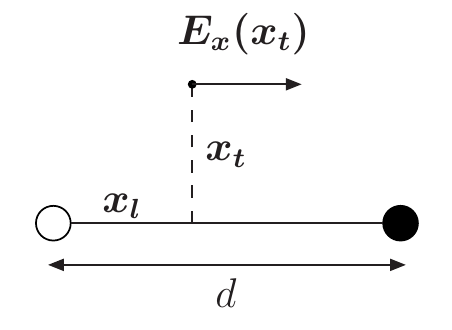}}
\caption{~\protect\subref{fig:operator_Wilson} The connected correlator given in Eq.~(\protect\ref{rhoW})
between the plaquette $U_{P}$ and the Wilson loop
(subtraction in $\rho_{W,\,\mu\nu}^{\rm conn}$ not explicitly drawn).
~\protect\subref{fig:qqbar} The longitudinal chromoelectric field $E_x(x_t)$ relative to the
position of the static sources (represented by the white and black circles),
for a given value of the transverse distance $x_t$.}
\label{fig:op_W}
\end{figure}

The field configurations generated by a static $q \bar q$ pair can be
probed by calculating on the lattice the vacuum expectation value of
the following connected correlation
function~\cite{DiGiacomo:1989yp,DiGiacomo:1990hc,Kuzmenko:2000bq,DiGiacomo:2000va}:
\begin{equation}
\label{rhoW}
\rho_{W,\,\mu\nu}^{\rm conn} = \frac{\left\langle {\rm tr}
\left( W L U_P L^{\dagger} \right)  \right\rangle}
              { \left\langle {\rm tr} (W) \right\rangle }
 - \frac{1}{N} \,
\frac{\left\langle {\rm tr} (U_P) {\rm tr} (W)  \right\rangle}
              { \left\langle {\rm tr} (W) \right\rangle } \; .
\end{equation}
Here $U_P=U_{\mu\nu}(x)$ is the plaquette in the $(\mu,\nu)$ plane, connected
to the Wilson loop $W$ by a Schwinger line $L$, and $N$ is the number of colors
(see Fig.~\ref{fig:op_W}).

The correlation function defined in Eq.~(\ref{rhoW}) measures the field 
strength $F_{\mu\nu}$, since in the naive continuum limit~\cite{DiGiacomo:1990hc}
\begin{equation}
\label{rhoWlimcont}
\rho_{W,\,\mu\nu}^{\rm conn}\stackrel{a \rightarrow 0}{\longrightarrow} a^2 g 
\left[ \left\langle
F_{\mu\nu}\right\rangle_{q\bar{q}} - \left\langle F_{\mu\nu}
\right\rangle_0 \right]  \;,
\end{equation}
where $\langle\quad\rangle_{q \bar q}$ denotes the average in the presence of 
a static $q \bar q$ pair, and $\langle\quad\rangle_0$ is the vacuum average.
This relation is a necessary consequence of the gauge-invariance of the
operator defined in Eq.~(\ref{rhoW}) and of its linear dependence on the
color field in the continuum limit (see Ref.~\cite{Cea:2015wjd}).

The lattice definition of the quark-antiquark field-strength tensor $F_{\mu\nu}$  is then obtained
by equating the two sides of Eq.~(\ref{rhoWlimcont}) for finite lattice spacing.
In the particular case when the Wilson loop $W$ lies in the plane
with $\hat \mu=\hat 4$ and $\hat \nu=\hat 1$ (see Fig.~\ref{fig:operator_Wilson})
and the plaquette $U_P$ is placed in the planes
$\hat 4\hat 1$, $\hat 4\hat 2$, $\hat 4\hat 3$, $\hat 2\hat 3$, $\hat 3\hat 1$,
$\hat 1\hat 2$, we get, respectively, the color field components
$E_x$, $E_y$, $E_z$, $B_x$, $B_y$, $B_z$, at the spatial point corresponding to the position of the center of the plaquette, up to a sign depending on the
orientation of the plaquette.
Because of the symmetry of Fig.~\ref{fig:op_W}, the color fields take on the same values at spatial points 
connected by rotations  around the axis on which the sources are located
(the $\hat 1$- or $x$-axis in the given example) .

As  far as the color structure of the field $F_{\mu \nu}$  is concerned, we note that the source of  $F_ {\mu \nu}$ is the Wilson loop connected to the  plaquette in Fig.~\ref{fig:op_W}.
The role of the Schwinger lines entering in Eq.~(\ref{rhoW}) is to realize the color parallel transport between the source loop and  the ``probe" plaquette. 
The Wilson loop defines a direction in color space. The color field, Eq.~(\ref{rhoWlimcont}),  that we measure points in a color direction parallel to this direction, 
the color direction of the source. (There are fluctuations of the color fields in the other color directions. These should contribute to the width of the energy density.) 

In principle, the operator in Eq.~(\ref{rhoW}) could be affected by
  $x_t$-dependent renormalization effects, related to Schwinger lines,
  which might  contaminate the $x_t$ dependence of the
  color fields. However,  our data satisfy continuum scaling; as carefully checked
  in Ref.~\cite{Cea:2017ocq}, fields obtained in the same {\em physical} setup,
  but at different values of beta, are in perfect agreement in the range of
  parameters used in the present work. This would have been impossible in the
  presence of sizeable renormalization effects. The absence of such effects
  is probably explained by the fact that we perform smearing before
  taking measurements (see below), and smearing effectively amounts to
  pushing the system towards the continuum, where renormalization effects
  become negligible.

  In~\ref{smearing} we present some further discussion
  about the smearing procedure and, in particular, compare it with the
  approach based on the explicit renormalization of the operator given
  in~(\ref{rhoW}), recently pursued in Ref.~\cite{Battelli:2019lkz}.

\section{Lattice setup}
  \label{sec:setup}

We performed all simulations in pure gauge SU(3), with the standard Wilson
action as the lattice discretization. A summary of the runs performed is given
in Table~\ref{tab:runs}.
The error analysis was performed by the jackknife method over bins at different
blocking levels.
%
\begin{table*}[t]
\begin{center} 
  \caption{Summary of the runs performed in the SU(3) pure gauge theory
  (measurements are taken every 100 upgrades of the lattice configuration).}
  \label{tab:runs}
\newcolumntype{Y}{>{\centering\arraybackslash}X}
\begin{tabularx}{0.9\linewidth}{@{}|Y|Y|Y|Y|Y|Y|Y|@{}}
\hline\hline
$\beta$ & lattice & $a$[fm] & $d$ [lattice] & $d$ [fm] & statistics & smearing steps \\ \hline
6.370  & $48^4$  & 0.059 & 16 & 0.951(11)  &  5300  & 100  \\
6.240  & $48^4$  & 0.071 & 16 & 1.142(15)  &  21000 & 100  \\
6.136  & $48^4$  & 0.083 & 16 & 1.332(20)  &  84000 & 120  \\
\hline\hline 
\end{tabularx} 
\end{center}
\end{table*}
We set the physical scale for the lattice spacing by using the value
$\sqrt{\sigma} = 420 ~$MeV for the string tension, and the parameterization~\cite{Edwards:1998xf}
for $a\sqrt {\sigma}$ that gave an accurate fit in a high-statistics simulation
for all $\beta$ in the range $5.6 \le \beta \le 6.5$. The correspondences between
$\beta$ and the distance $d$ shown in Table~\ref{tab:runs} were obtained from this parameterization. Note that the distance in lattice units between quark and antiquark, corresponding to the size of the Wilson loop in the connected correlator in Eq.~\eqref{rhoW}, was kept fixed to $d=16\,a$.

The connected correlator defined in Eq.~(\ref{rhoW}) suffers from large
fluctuations at the scale of the lattice spacing, which are responsible
for a bad signal-to-noise ratio. To extract the physical information carried
by fluctuations at the physical scale (and, therefore, at large distances
in lattice units) we smoothed out configurations by a {\em smearing}
procedure. Our setup consisted of (just) one step of HYP
smearing~\cite{Hasenfratz:2001hp} on the temporal links, with smearing
parameters $(\alpha_1,\alpha_2,\alpha_3) = (1.0, 0.5, 0.5)$, and
$N_{\rm APE}$ steps of APE smearing~\cite{Falcioni1985624} on the spatial links,
with smearing parameter $\alpha_{\rm APE} = 0.25$. Here $\alpha_{\rm APE}$ is the
ratio of the weight of one staple to the weight of the original link.

\section{Numerical results}

\begin{figure}[tp]
   \centering
   \subfigure[$E_x(x_t,x_l)$]%
             {\label{fig:FieldsEx}\includegraphics[width=0.45\textwidth,clip]{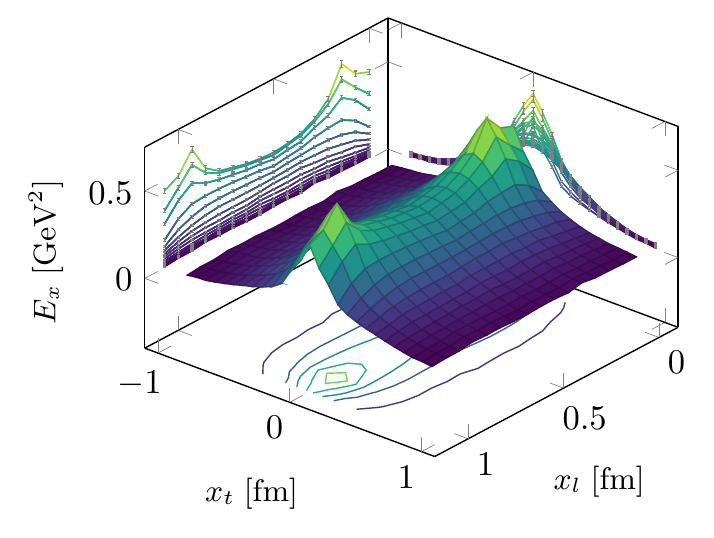}}\vfill
   \subfigure[$E_y(x_t,x_l)$]%
             {\label{fig:FieldsEy}\includegraphics[width=0.45\textwidth,clip]{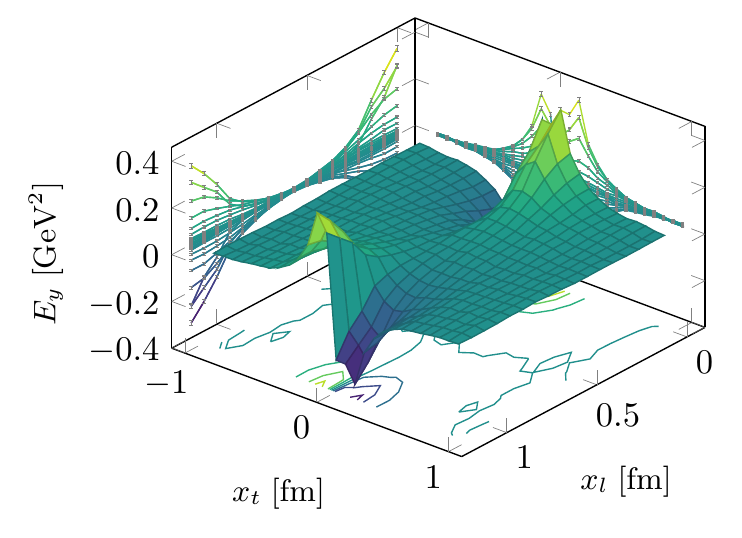}}\vfill
   \subfigure[$E_z(x_t,x_l)$]%
             {\label{fig:FieldsEz}\includegraphics[width=0.45\textwidth,clip]{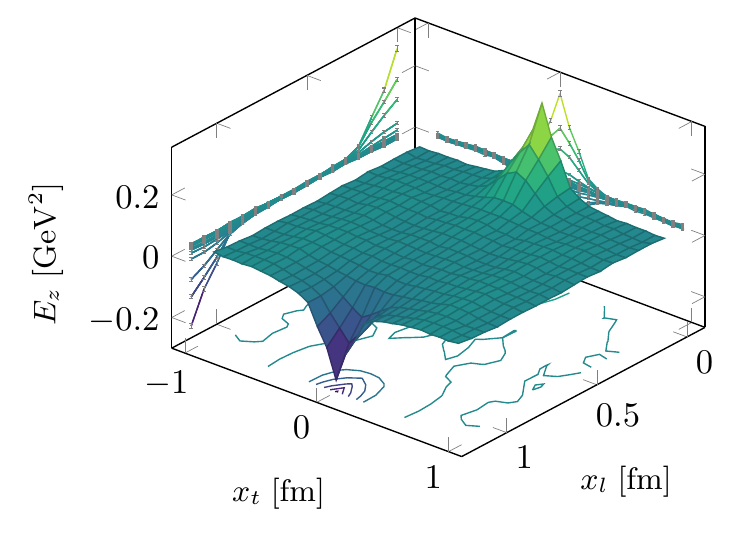}}
   \caption{Surface and contour plots for the three components of the
chromoelectric field at $\beta=6.240$ and $d=1.142$ fm.
All plotted quantities are in physical units.}
   \label{fig:FieldsE}
\end{figure}
\begin{figure}[tp]
   \centering
   \subfigure[$B_x(x_t,x_l)$]%
             {\label{fig:FieldsEx}\includegraphics[width=0.45\textwidth,clip]{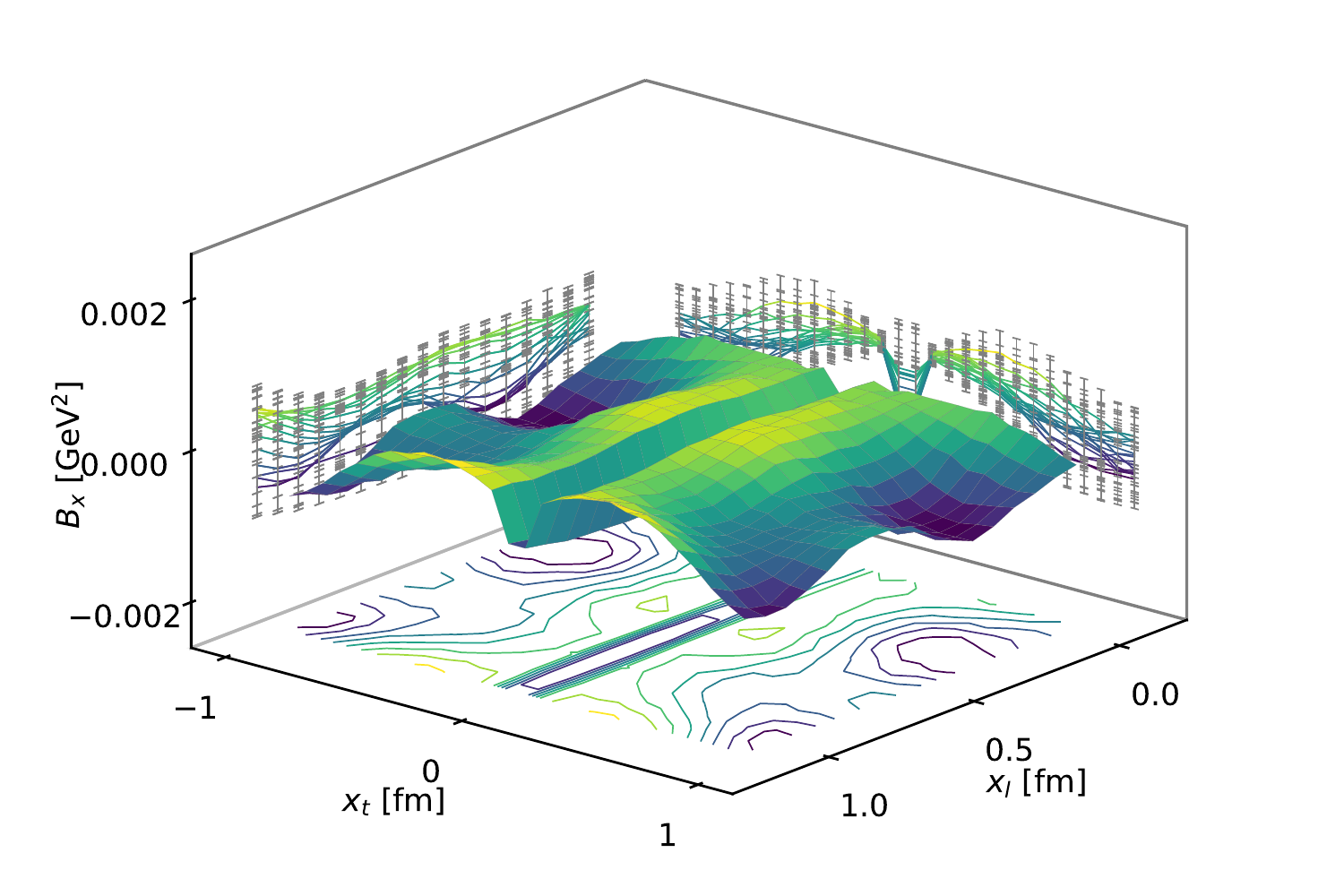}}\vfill
   \subfigure[$B_y(x_t,x_l)$]%
             {\label{fig:FieldsEy}\includegraphics[width=0.45\textwidth,clip]{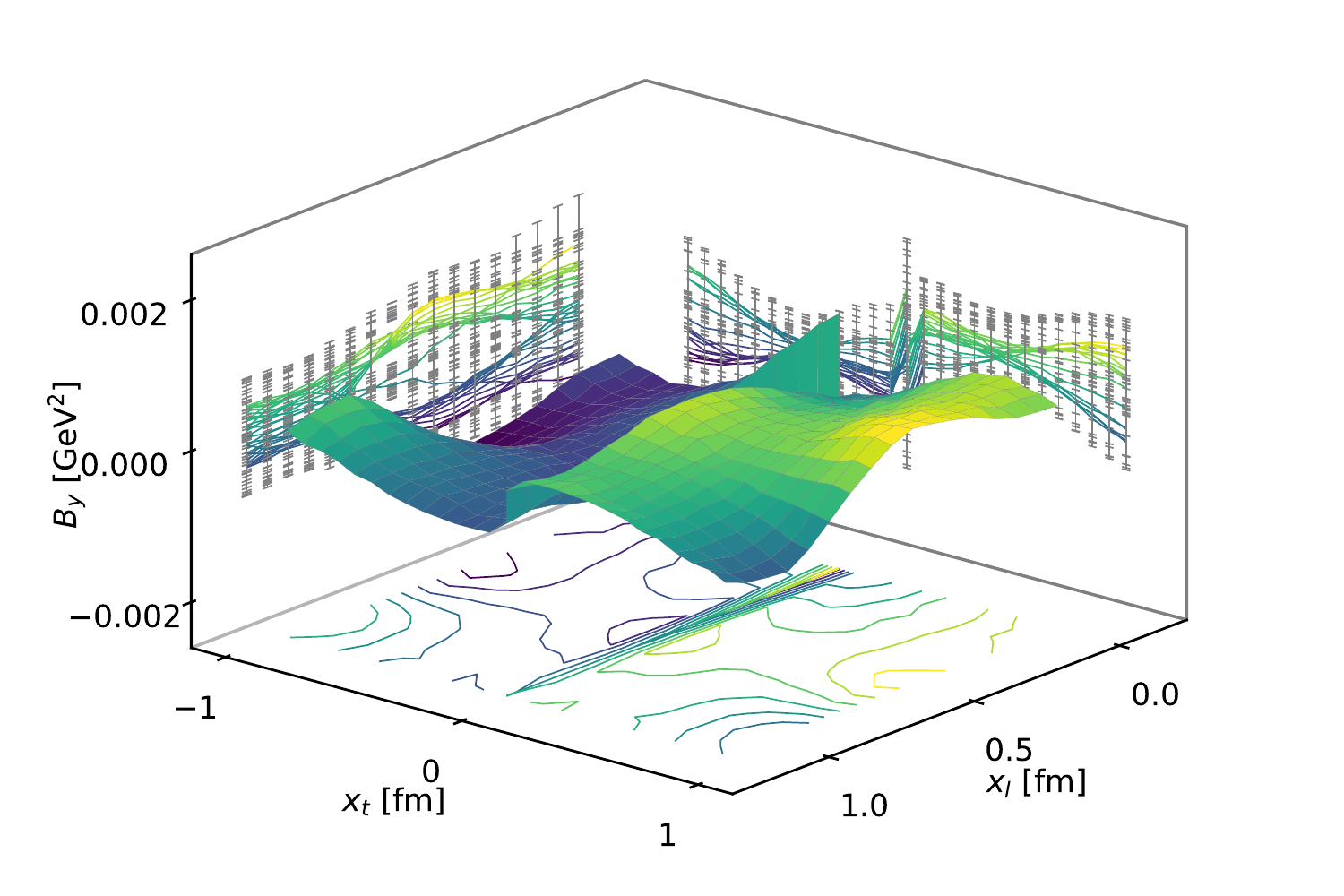}}\vfill
   \subfigure[$B_z(x_t,x_l)$]%
             {\label{fig:FieldsEz}\includegraphics[width=0.45\textwidth,clip]{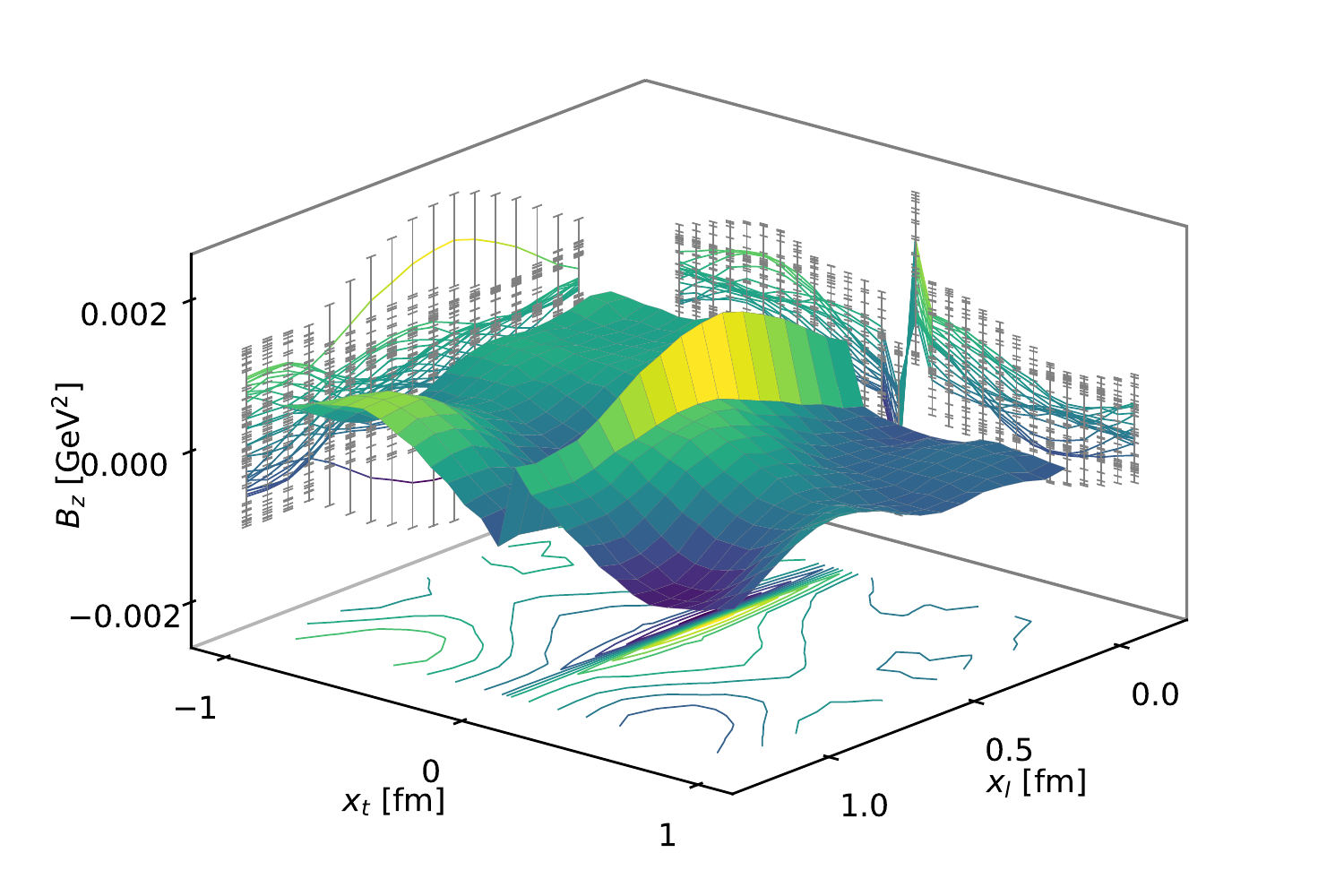}}
   \caption{Surface and contour plots for the three components of the
chromomagnetic field at $\beta=6.240$ and $d=1.142$ fm.
All plotted quantities are in physical units.}
   \label{fig:FieldsB}
\end{figure}

Using Monte Carlo evaluations of the expectation value of the operator
$\rho_{W,\,\mu\nu}^{\rm conn}$ over smeared ensembles, we have determined the six components of
the color fields on all  two-dimensional planes transverse to the
line joining the color sources  allowed by the lattice discretization.
These measurements were carried out for three values of the
distance $d$ between the
static sources, at values of $\beta$ lying inside the
continuum scaling region, as determined in Ref.~\cite{Cea:2017ocq}.

We found that the chromomagnetic field is everywhere much smaller than the
longitudinal chromoelectric field and is compatible with zero within
statistical errors (see Fig.~\ref{fig:FieldsB}). As expected, the dominant
component of the chromoelectric field is longitudinal, as is seen in 
Fig.~\ref{fig:FieldsE}, where we plot  the components of the simulated
chromoelectric field $\vec{E}$ at $\beta = 6.240$ as functions of  their longitudinal  displacement
from one of the quarks,  $x_l$, and  their transverse distance from the axis, $x_t$.

While  the transverse components of the chromoelectric field are also smaller
than the longitudinal component, they are larger than the statistical errors in a
region wide enough that we can match them  to the transverse components of an effective
Coulomb-like field $\vec{E}^C (\vec{r})$  produced by two static sources. 
For points which are not very close to the quarks,
this matching can be carried out with a single fitting parameter $Q$,  the effective charge of 
static quark and antiquark  sources determining   $\vec{E}^C (\vec{r})$. %

To the extent that we can fit the  transverse components of the simulated field
$\vec{E}$ to  those of $\vec{E}^C (\vec{r})$ with an appropriate choice of $Q$, the
nonperturbative  difference  $\vec{E}^{NP}$  between the simulated chromoelectric
field $\vec{E} $ and the effective Coulomb field $\vec{E}^C$
\begin {equation}
\vec{E}^{NP} ~\equiv~\vec{E}~-~\vec{E}^C.  
\label{ENP}
\end{equation}
will be purely  longitudinal.  We then identify $\vec{E}^{NP}$ as the confining field
of the QCD flux tube.

\section{Evaluation of the effective Coulomb field of the sources}

To extract  the longitudinal component of the confining field  $\vec{E}^{NP}$,
Eq.~(\ref{ENP}), from lattice simulations,  we must first determine the
effective charge  of the sources, $Q$, by fitting the transverse components of
the simulated field to those of an effective Coulomb field
$\vec{E}^C (\vec{r})$:
%
\begin{eqnarray}
\label{C2}
& \vec{E}^C (\vec{r}) \; = \;  Q \left( \frac{\vec{r}_1}{\max(r_1, R_0)^3} \;
- \;  \frac{\vec{r}_2}{\max(r_2, R_0)^3} \; \right ) \; ,
\\ \nonumber
& \vec{r}_1\equiv \vec r -\vec r_Q\;, \;\;\;
\vec{r}_2\equiv\vec r - \vec r_{-Q}\;,
%
\end{eqnarray}
where $\vec r_Q$ and $\vec r_{-Q}$ are the positions of the two static color
sources and $R_0$ is the effective radius of the color source, introduced to explain,
at least partially,  the decrease of the field close to the sources.
Due to the axial symmetry around the line connecting the static charges\footnote{We
have explicitly checked that  within statistical
errors the color field distributions respect  this axial symmetry.}, we may consider
the color field distributions in the $x\,y$ plane without loss of generality. Then $x \equiv x_l,~y \equiv x_t$.

We find that with an appropriate choice of $Q$ the $y$-com\-ponent  of the simulated
chromoelectric field, $E_y$,  is approximately equal to the $y$-component of the
Coulomb field, $E_y^C$
 at distances greater than 1--2 lattice spacings from the quarks.
In making the fit we must take into account that the color fields are probed by
a plaquette, so that the measured field value should be assigned to the center of the
plaquette. This also means that the $z$-component of the field is  probed at 
a distance of $1/2$ lattice spacing from the $x\,y$ plane, where the $z$-component 
of the Coulomb field $E_z^C$ is nonzero and can be matched with the measured value
$E_z$ for the same value of $Q$.

In Table~\ref{tab:coulombFit}, we list the values of the effective charge $Q$ obtained from lattice
measurements of $E_z$ and $E_y$ at three values of $d$, the
quark-antiquark separation.
%
\begin{table*}[t]
\begin{center}
\caption{Values of the fit parameters $Q$ and $R_0$ extracted from Coulomb fits of the transverse
components of the chromoelectric field 
and values of the longitudinal chromoelectric fields 
at $(d/2,0)$, the midpoint 
between the sources and transverse distance zero,
for several values of distance $d$.
$E_x(d/2,0)$  
is the unsubtracted simulated field and $E_{x}^{NP}(d/2,0)$
is the nonperturbative
chromoelectric field.
For the parameters of the Coulomb fit we quote, along with the statistical error, a systematic uncertainty that accounts for the variability in the values of the fit parameters extracted from all acceptable fits to $E_{y}$ and $E_{z}$ at different $x_l$ values (for more details, see ~\ref{fit}).
}
\label{tab:coulombFit}
\newcolumntype{Y}{>{\centering\arraybackslash}X}
\begin{tabularx}{0.9\linewidth}{@{}|Y|Y|Y|Y|Y|Y|Y|@{}}
\hline\hline
$\beta$ & $d$ [fm] & $Q$ & $R_0$ [fm] & $E_x(d/2,0)$ $[\text{GeV}^2$] & $E_{x}^{NP}(d/2,0)$  $[\text{GeV}^2]$ \\ \hline
6.370  & 0.951(11) & 0.278(4)(43)  &  0.1142(16)(200) &  0.360(9)  & 0.263(7) \\
6.240  & 1.142(15) & 0.289(11)(38) &  0.1367(29)(241) & 0.335(11) & 0.265(10) \\
6.136  & 1.332(20) & 0.305(14)(81) &  0.179(6)(32)     & 0.288(25) & 0.234(25) \\
\hline\hline
\end{tabularx}.
\end{center}
\end{table*}
The statistical uncertainties in the quoted $Q$ values result from the comparisons
among Coulomb fits of $E_y$ and $E_z$ at the values of $x_l$, 
for which we were able to get meaningful results for the fit. The
stability of $Q$ under a change of the fitting strategy, its dependence
upon the values of $x_l$ included in the fit and the global assessment of the
systematic uncertainties will be presented in a forthcoming extended version of
this work. The values of $R_0$ in physical units grow with the growth 
of the lattice step $a$, while in lattice units they show more stability.
This suggests that the effective size of a color charge in our case 
is mainly explained by lattice discretization artifacts and the smearing
procedure, and is not a physical quantity.
In~\ref{fit} we present some details about the Coulomb fit.

Evaluating the contribution of the field of the quark to $\bold{E}^C(\bold{r})$ in Eq.~(4) at the position $\bold{r_{-Q}}$ of the antiquark and multiplying by the charge $- 4 \pi Q$ of the antiquark yields a Coulomb force between the quark and antiquark with coefficient $- 4 \pi Q^2$.  By comparison, in the standard string picture of the color flux tube, a Coulomb correction  of strength $-\pi/12$ to the long distance linear potential  (the universal L\"uscher term) arises from the long wave length transverse fluctuations of the flux tube~\cite{Luscher:1980ac}. This L\"uscher term is equal to the Coulomb force generated by the field $\bold{E}^C(\bold{r})$ (4) when $Q = 1/(4 \sqrt{3}) \approx 0.144$. This is roughly $1/2$ the values of Q measured in our simulations and listed in Table 2. (We note the L\"uscher value of the Coulomb force is consistent with the results~\cite{Necco:2001xg,Kaczmarek:2005ui,Karbstein:2018mzo} of lattice situations of the heavy quark potential at distances down to $\approx 0.4$ fm.) 
Although the connection between these two descriptions of the Coulomb force is not clear, we note that the color fields we measure point in a single direction in color space. The fluctuating color fields in the other color directions might affect the strength of the effective Coulomb force between the quark and the antiquark.

\section{Evaluation of the nonperturbative color field}

\begin{figure}
   \centering
   \subfigure[$E_x^{NP}(x_t,x_l)$]%
             {\label{fig:Fields_confiningEx}\includegraphics[width=0.45\textwidth,clip]{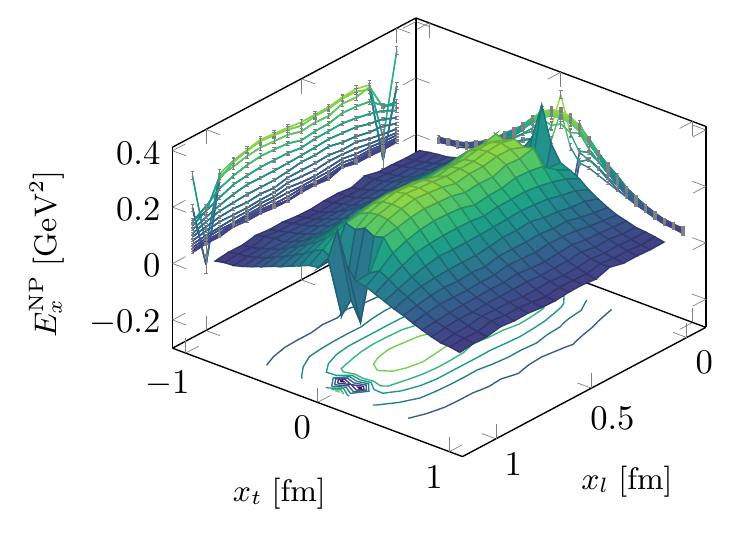}}\vfill
   \subfigure[$E_y^{NP}(x_t,x_l)$]%
             {\label{fig:Fields_confiningEy}\includegraphics[width=0.45\textwidth,clip]{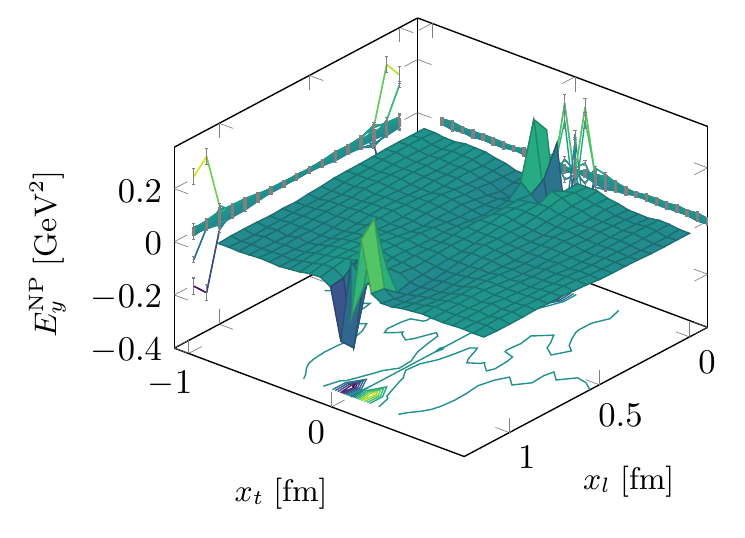}}\vfill
   \subfigure[$E_z^{NP}(x_t,x_l)$]%
             {\label{fig:Fields_confiningEz}\includegraphics[width=0.45\textwidth,clip]{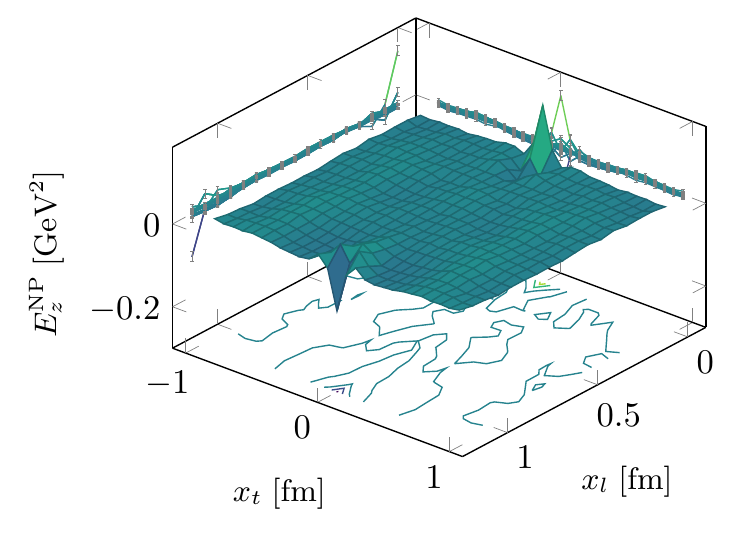}}
\caption{Surface and contour plots for the three components of the nonperturbative
chromoelectric field, $\vec{E}^{NP} \equiv \vec{E} - \vec{E}^C$, at $\beta = 6.240$ and
$d = 1.142~ $fm. All plotted quantities are in physical units.}
\label{fig:Fields_confining}
\end{figure}

Once $\vec{E}^C$  has been fixed, the difference Eq.~\eqref{ENP} between the simulated
field $\vec{E}$ and the field $\vec{E}^C$ determines $\vec{E}^{NP}$. In this way we
obtain the {\em nonperturbative} structure of the flux tube. To the best of
our knowledge, this is the first time that a confining part of the measured
longitudinal chromoelectric field has been extracted making use only of lattice
data.

In Fig.~\ref{fig:Fields_confining} we plot the longitudinal component  $E_x^{NP}$ of the 
nonperturbative field (3)
as a function of the longitudinal and
transverse displacements $x_l$, $x_t$ at $\beta = 6.240$. As expected,
$E^{NP}$ is almost uniform along the flux tube
at  distances not too close to the static color sources.
This feature is better seen in Fig.~\ref{fig:NPEx},
where  transverse  sections of the  field  $E_x^{NP} ( x_l, x_t)$, plotted in
Fig.~\ref{fig:Fields_confining}, are shown for the values of $x_l$ specified in
Fig.~\ref{fig:NPEx}. For these values of $x_l$  the shape
of the nonperturbative longitudinal field is basically constant all along the
axis. 
\begin{figure}[htb] 
\centering
\includegraphics[width=1.0\linewidth,clip]{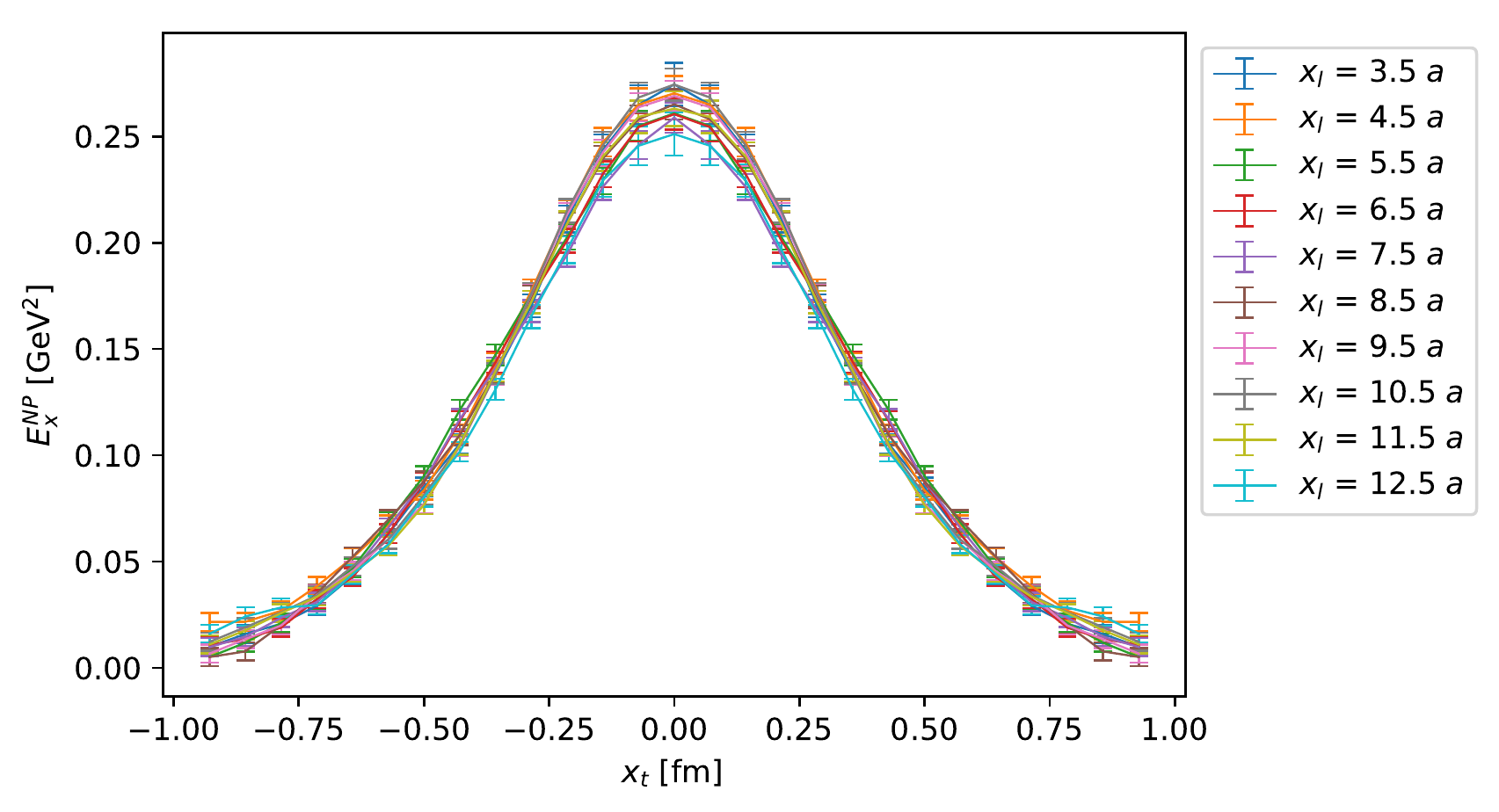}
\caption{Transverse cross sections of the nonperturbative field
$E_x^{NP}(x_t)$ at $\beta=6.240$, $d=1.142$ fm, for several values of $x_l$.
}
\label{fig:NPEx}
\end{figure}
Although figures 2,3, and 4 refer only to the case of
$\beta = 6.240$ and $d=1.142$ fm, the scenario is similar for the
other two lattice setups listed in Table~\ref{tab:runs}.

In Table~\ref{tab:coulombFit} we also compare the values of the measured longitudinal
chromoelectric field $E_x$ with those of  the nonperturbative field
$E_x^{NP}$ on the axis at the midpoint between the quark and antiquark,
for all three values of their separation $d$. Given that   $E^{NP}_x$ is almost
uniform along the axis, $E_x^{NP} (x_l, x_t=0)$ assumes these same values
at all points $x_l$ on the axis for all distances larger than approximately
$0.1-0.2 ~$fm from the quark sources.The value of  $E_x^{NP} (x_l, x_t=0)$ is 
closely related to the value of the string tension (see below). 
In future work we will use the distribution of color fields
between a quark and an antiquark to calculate the force between them.

\section{Future work}

The value of the chromoelectric field at the position of the quarks is equal to force on the quarks~\cite{Brambilla:2000gk}, 
{\it i.e.} , the derivative of the heavy quark potential. 
However because the difficulty of carrying accurate simulations of the color field close to position of the source 
we cannot use our  current simulations to  determine  the  quark-antiquark force as a function of their separation.  
This is one goal of our future work. 

We plan to compare the stress tensor density distribution calculated from our simulated color fields  with the stress tensor density simulated in Ref.~\cite{Yanagihara:2018qqg}. 
This will provide a test of our vision of the space between static color charges as filled with lines of force of chromoelectric fields pointing in a color direction parallel 
to the color direction of the source, generating a Maxwell-like stress tensor in this color direction. 
Since there are fluctuations of the color fields in the other color directions, the width of the stress tensor density that we measure should be interpreted as  the intrinsic width of the flux tube.

We can use our simulated stress tensor to calculate the force transmitted across the midplane and the resulting quark-antiquark force for the values of their separations where our simulations are carried out. These predictions  can then be tested  by comparing our simulated results  with a parameterization of Wilson loop data for the long distance heavy quark potential.   To the extent that the nonperturbative field $E^{NP}$ generates the long distance constant heavy quark force ( i.e., the string tension),  and that the Coulomb-like field $E^C$ generates the $1/R^2$ correction to it,
 the force between a quark and an antiquark can be understood in terms of the color fields permeating the space between them.

\section{Summary}

In this paper we have determined the spatial distribution in three dimensions of
all components of the color fields generated by a static quark-antiquark pair. 
We have found that the dominant component of the color field is the
chromoelectric one in the longitudinal direction, {\it i.e.} in the
direction along the axis connecting the two quark sources. This feature
of the field distribution has been known for a long time. However, the
accuracy of our numerical results allowed us to go far
beyond this observation. First, we have found that all the chromomagnetic
components of the color field are compatible with zero within the statistical
uncertainties. Second, the chromoelectric components of the color fields
in the directions transverse to the axis connecting the two sources,
though strongly suppressed with respect to the longitudinal component,
are sufficiently greater than the statistical uncertainties that we
could manage to interpolate them.

 Our remarkable finding was that the transverse components of the simulated
chromoelectric field can be nicely reproduced  by a Coulomb-like field generated
by two  sources with opposite charge (everywhere except in a small region around
the sources). We then subtract this Coulomb-like field from the simulated
chromoelectric field to obtain a nonperturbative field $\vec{E}^{NP}$ according
to Eq.~\eqref{ENP}.
The dependence of the resulting longitudinal component of  $\vec{E}^{NP}$  on the distance
$x_t$ from the axis is independent of the position $x_l$ along the
axis, except near the sources.
We identify the nonperturbative field found in this way from lattice simulations
as the confining field of the QCD flux tube.

\section{Discussion}

We stress that our separation of the chromoelectric field into perturbative and
nonperturbative components was obtained by directly analyzing lattice data on
color field distributions between static quark sources. To the best of our
knowledge this separation between perturbative and nonperturbative
components has not been carried out previously.

The idea of this separation is independent of the procedure used in this paper to implement it.  The separation provides a new tool with which to
probe the chromoelectric field surrounding the quarks. Our approach can be
straightforwardly extended to the case of QCD with dynamical fer\-mions with physical
masses and at nonzero temperature and baryon density.

\section*{Acknowledgements}
This investigation was in part based on the MILC collaboration's public
lattice gauge theory code. See {\url{http://physics.utah.edu/~detar/milc.html}}.
Numerical calculations have been made possible through a CINECA-INFN
agreement, providing access to resources on MARCONI at CINECA.
AP, LC, PC, VC acknowledge support from INFN/NPQCD project.
FC acknowledges support from the German Bundesministerium f{\"u}r Bildung und
Forschung (BMBF) under Contract No. 05P1RFCA1/05P2015 and from the DFG
(Em\-my Noether Programme EN 1064/2-1).
VC acknowledges financial support from the INFN HPC{\_}HTC project.

\appendix

\section{Smearing and renormalization}
\label{smearing}

\begin{figure}[ht] 
\centering
\includegraphics[width=0.9\columnwidth,clip]{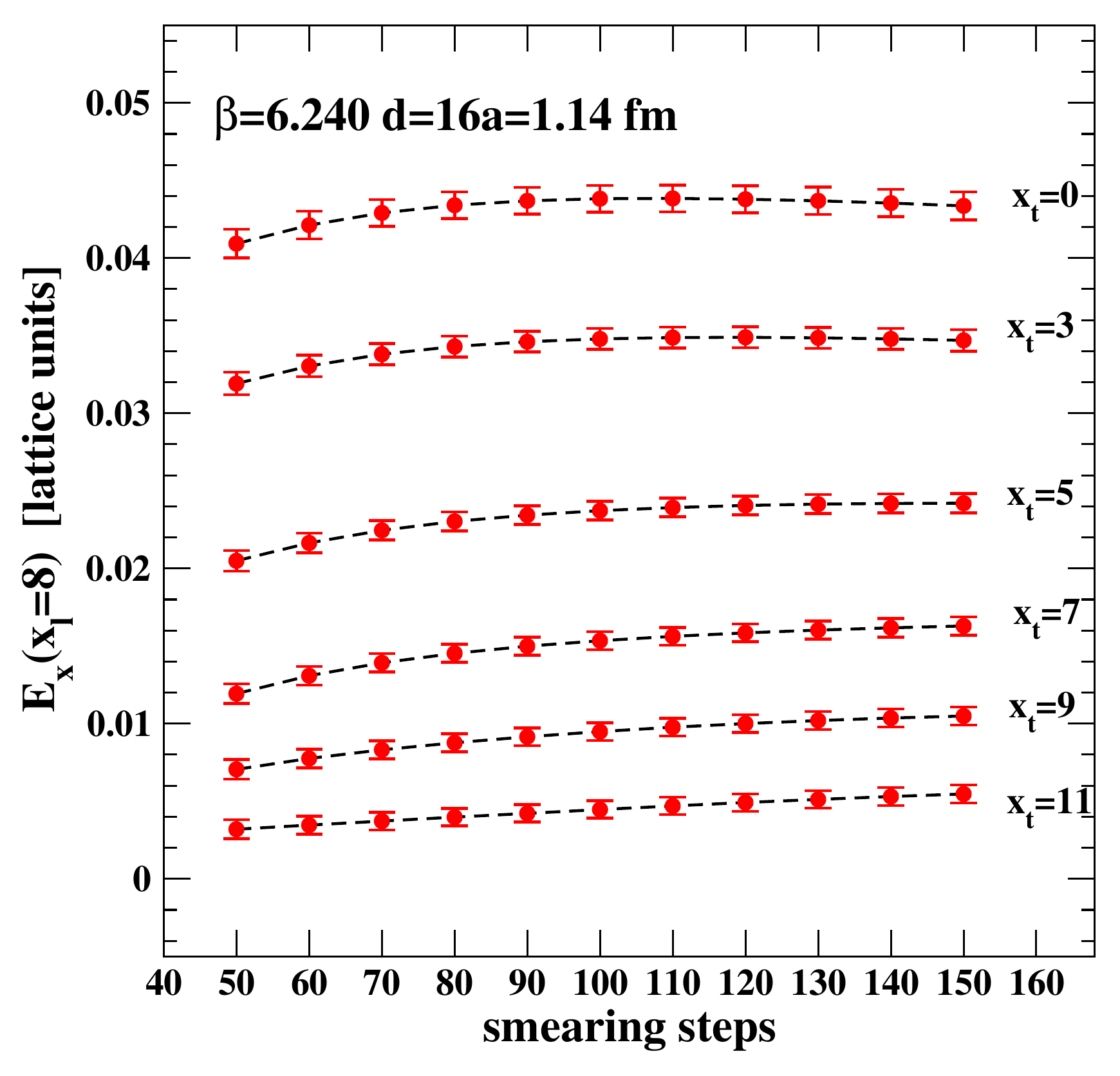}
\caption{$E_x$ in lattice units versus smearing at $\beta=6.240$, $d=1.14$ fm, measured at the midpoint between sources ($x_l=8a$), 
for several values of $x_t$ (in lattice units).}
\label{fig:NPEx}
\end{figure}
\begin{figure}[htb] 
\centering
\includegraphics[width=0.9\columnwidth,clip]{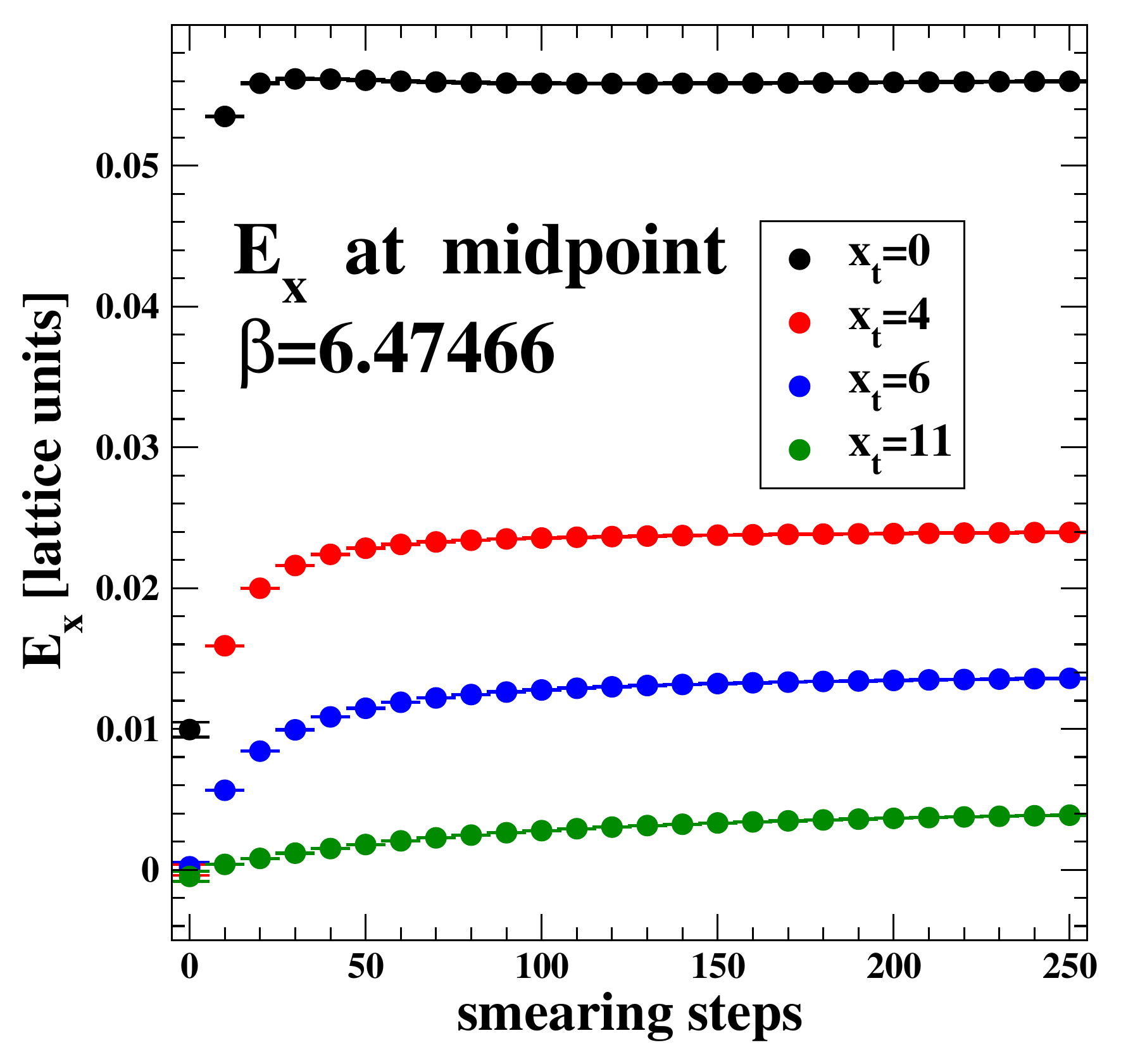}
\caption{$E_x$ in lattice units at several values of $x_t$ versus smearing
at $\beta=6.47466$ and $d=4a\simeq0.37$ fm as in Ref.~\cite{Battelli:2019lkz}.}
\label{fig:Ex_midpoint_vs_smearing}
\end{figure}
\begin{figure}[htb] 
\centering
\includegraphics[width=0.9\columnwidth,clip]{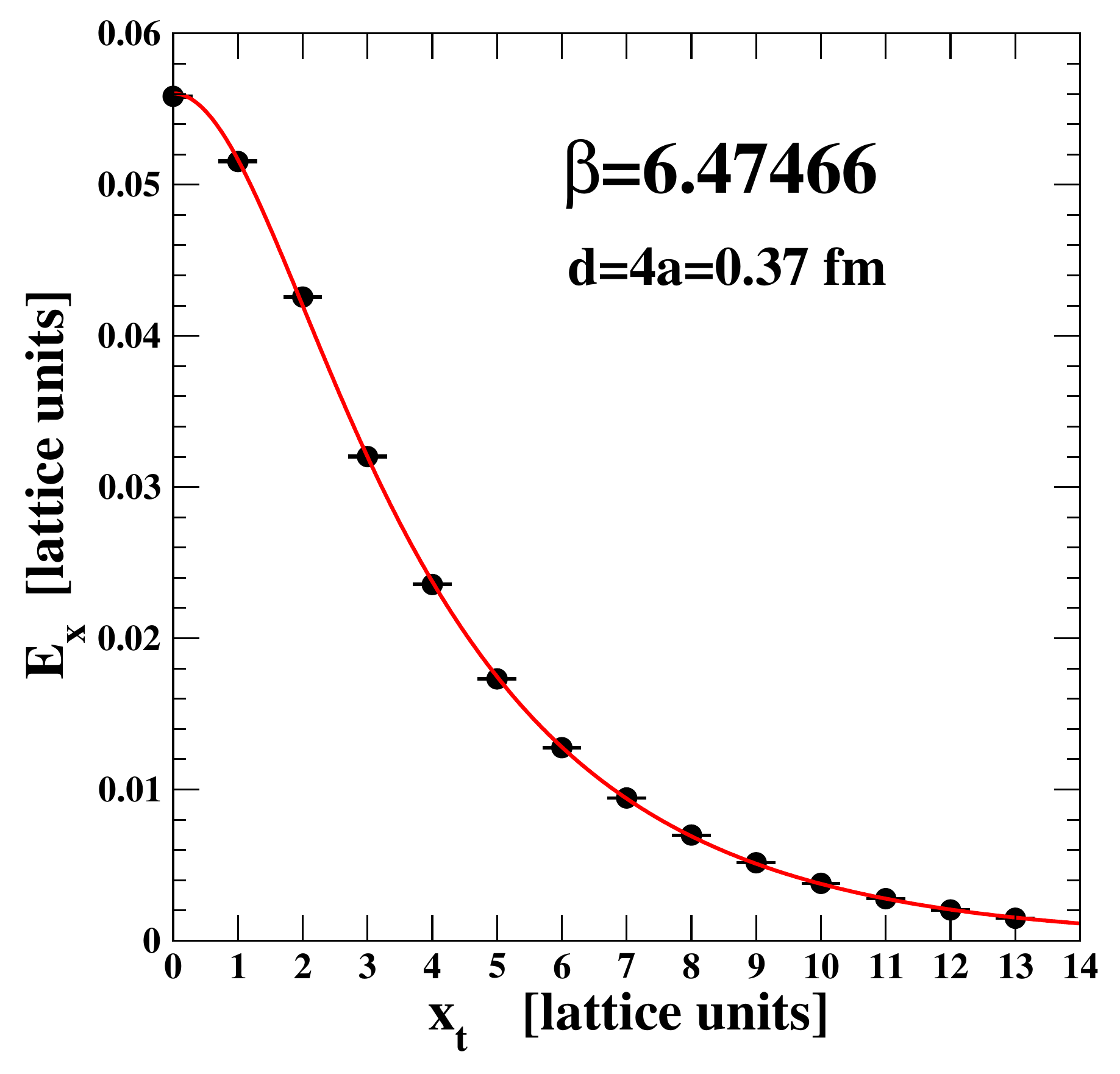}
\caption{Clem fit (see Eq.~924) of Ref.~\cite{Battelli:2019lkz} to $E_x$ in
  lattice units at $\beta=6.47466$ and $d=4a\simeq0.37$~fm. The values of
  $E_x$ have been measured after 100 smearing steps.}
\label{fig:Clem_fit_to_Ex}
\end{figure}

  The typical behavior of the (unsubtracted) chromoelectric
  field in the longitudinal direction with our smearing setup is described in
  Fig.~\ref{fig:NPEx}, which gives, at $\beta=6.240$ and for a distance
  $d=1.14$~fm between the sources, the field $E_x$ measured at the midpoint
  between sources ($x_l=8a$) for several values of the transverse
  distance $x_t$. We can clearly see that, for increasing number of the smearing
  steps, $E_x$ reaches a plateau for larger and larger values of $x_t$, with
  no sign whatsoever of degradation of the signal. The value of the smearing
  step quoted in the last column of Table~\ref{tab:runs} is such that
  all field components, on all transverse planes and at all values of $x_t$
  (except possibly the few largest ones) have reached their plateaux.

\begin{figure}[htb] 
\centering
\includegraphics[width=0.9\columnwidth,clip]{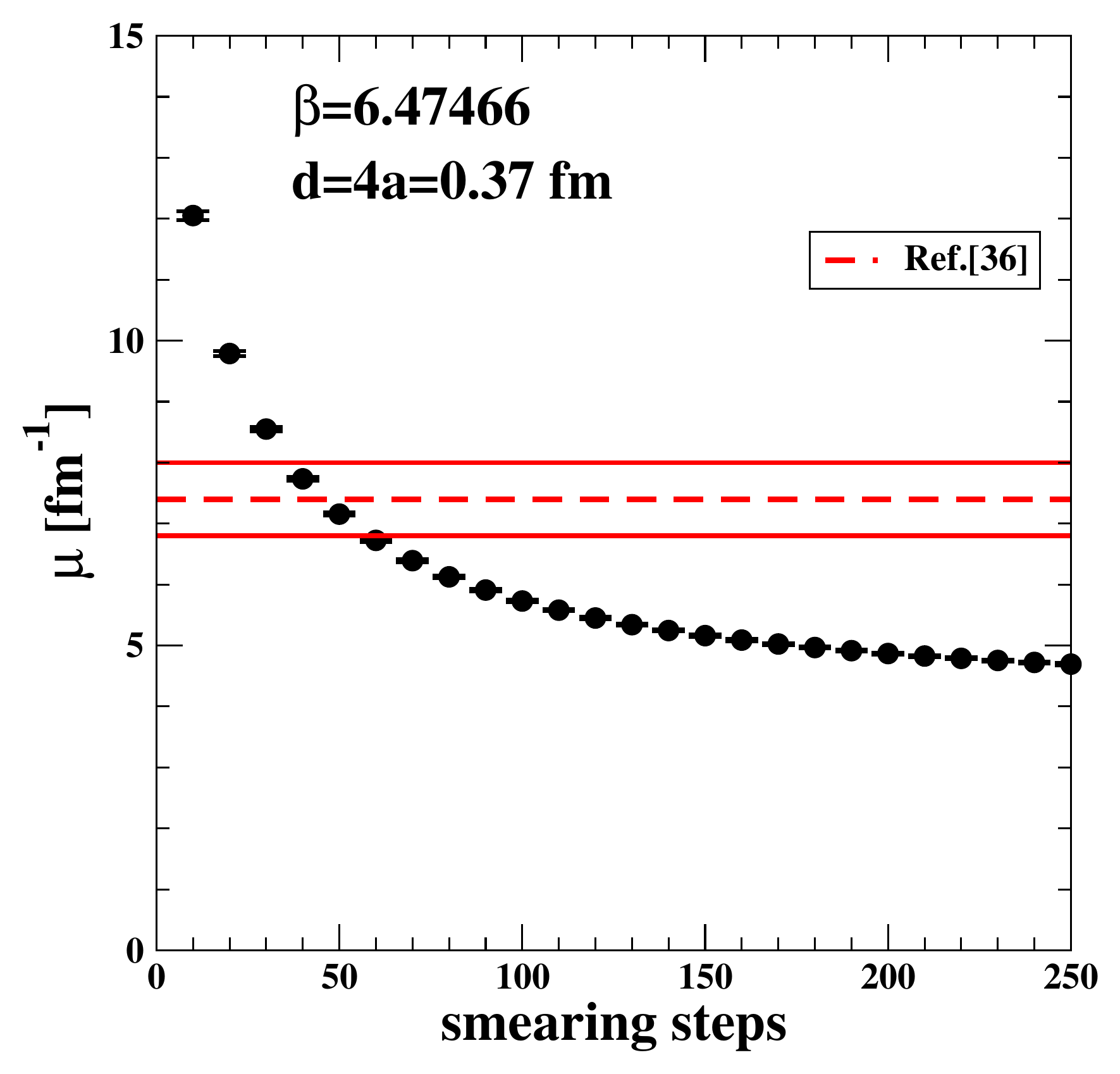}
\caption{The Clem $\mu$ parameter (full black circles) versus smearing  at $\beta=6.47466$ and
  $d=4a\simeq0.37$~fm compared to the value obtained in
  Ref.~\cite{Battelli:2019lkz} (the red dashed line is the central value, the red full lines delimit
  the error band).}
\label{fig:mu_midpoint_vs_smearing}
\end{figure}
\begin{figure}[htb] 
\centering
\includegraphics[width=0.9\columnwidth,clip]{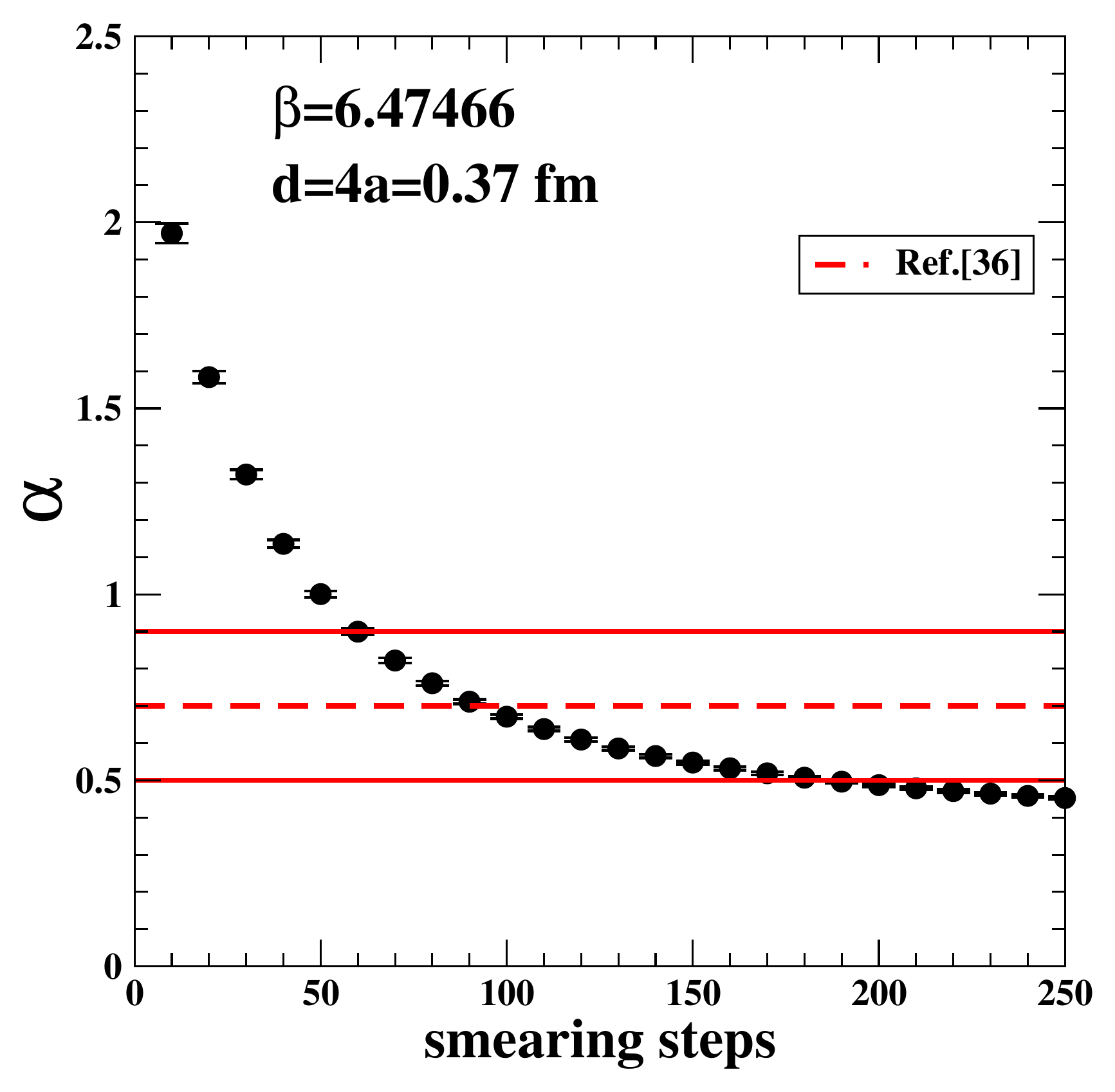}
\caption{The Clem $\alpha$ parameter (full black circles) versus smearing at $\beta=6.47466$ and
  $d=4a\simeq0.37$~fm compared to the value obtained in
  Ref.~\cite{Battelli:2019lkz} (the red dashed line is the central value, the red full lines delimit
  the error band).}
\label{fig:alpha_midpoint_vs_smearing}
\end{figure}
\begin{figure}[htb] 
\centering
\includegraphics[width=0.9\columnwidth,clip]{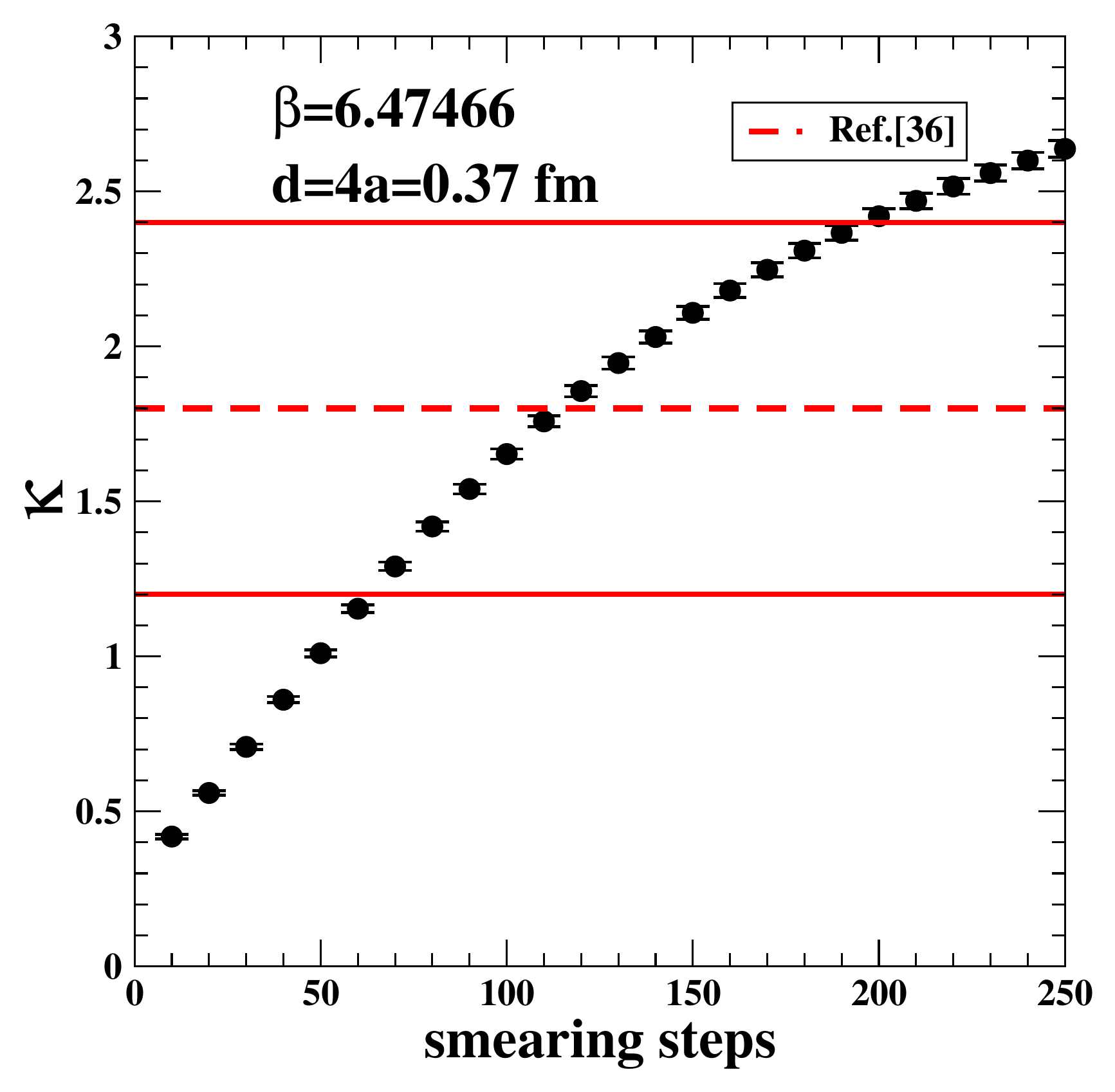}
\caption{The Clem $\kappa$ parameter (full black circles)  versus smearing at $\beta=6.47466$ and
  $d=4a\simeq0.37$~fm compared to the value obtained in
  Ref.~\cite{Battelli:2019lkz} (the red dashed line is the central value, the red full lines delimit
  the error band).}
\label{fig:kappa_midpoint_vs_smearing}
\end{figure}

Recently, a paper appeared~\cite{Battelli:2019lkz} which
  studied the flux tube between two static sources by means of a connected
  operator similar to ours, except that the role of the Wilson loop is
  replaced by two parallel and oppositely directed Polyakov loops. No 
smoothing  is performed on the ensemble configuration, but 
  the renormalization properties of the connected operator are properly
  taken into account. Their analysis of the (unsubtracted) longitudinal
  chromoelectric field at the midpoint between two sources  separated by  distance
  $d=4a\simeq 0.37$~fm gave  the following
  values for the "Clem parameters''~\cite{Clem:1975aa} describing the
  transverse profile of the field:
  \begin {equation}
    \frac{1}{\lambda}=\mu=7.4(6) \ {\rm fm}^{-1}\;\;,\;\; \alpha=0.7(2)\ \;\;,
    \;\; \kappa=1.8(6) \,.
    \label{ClemparametersPisa}
  \end{equation}
(See Eq.(24) of  Ref.~\cite{Battelli:2019lkz}).

To compare the analysis Ref.~\cite{Battelli:2019lkz}  with our smearing procedure, we measured the longitudinal
  chromoelectric field $E_x$ at $\beta=6.47466$ and at a distance between the  
  two static sources equal to $d=4a$.  With the 
scale setting procedure
  used in Ref.~\cite{Battelli:2019lkz} this  distance corresponds in physical units to the same quark-antiquark separation $d = 0.37$ fm considered there. 
 The behavior under smearing of the measured (unsubtracted) field $E_x$ at the midpoint between the sources and at different values of $x_t$ 
 is shown in Fig.~\ref{fig:Ex_midpoint_vs_smearing}. The qualitative
  behavior is the same as in Fig.~\ref{fig:NPEx}: the larger the number of
  smearing steps, the greater the number of values of $x_t$ for which $E_x$ has reached
  a plateau. We use the Clem parameterization to fit the transverse profiles.  After 100 smearing steps, our determination of the parameters
  of the Clem fit gives
\begin {equation}
  \frac{1}{\lambda}=\mu=5.73(2) \ {\rm fm}^{-1}\;,\;\; \alpha=0.67(1)\;,\;\;
  \kappa=1.65(2) \,,
\label{Clemparameters}
\end{equation}
The values of $\alpha$ and $\kappa$ are in nice agreement with Ref.~\cite{Battelli:2019lkz} but the difference in the values of $\mu$ requires further investigation.  In Fig.~\ref{fig:Clem_fit_to_Ex} the fit is compared to
data: it seems to be qualitatively very good,
in spite of a $\chi^2/{\rm dof}$ of about $40$, probably due to the large
correlation among data at different $x_t$, which were obtained in the same
Monte Carlo simulation.

In Figs.~\ref{fig:mu_midpoint_vs_smearing}, \ref{fig:alpha_midpoint_vs_smearing}
and \ref{fig:kappa_midpoint_vs_smearing}, we show the behavior of the
parameters of the Clem fit versus smearing at $\beta=6.47466$ and
$d=4a\simeq0.37$~fm and compare them with the corresponding values as quoted
in Ref.~\cite{Battelli:2019lkz}. The conclusion that can be drawn is that
smearing behaves, not surprisingly, as an effective renormalization, driving
the parameters towards the values extracted from the renormalized field. Of
course, the different systematics in the two approaches must be carefully
studied, to improve the matching.

\section{Coulomb fit}
\label{fit}

\begin{figure}[htb] 
\centering
   \subfigure[]%
             {\label{fig:CoulombEy}\includegraphics[width=0.9\columnwidth,clip]{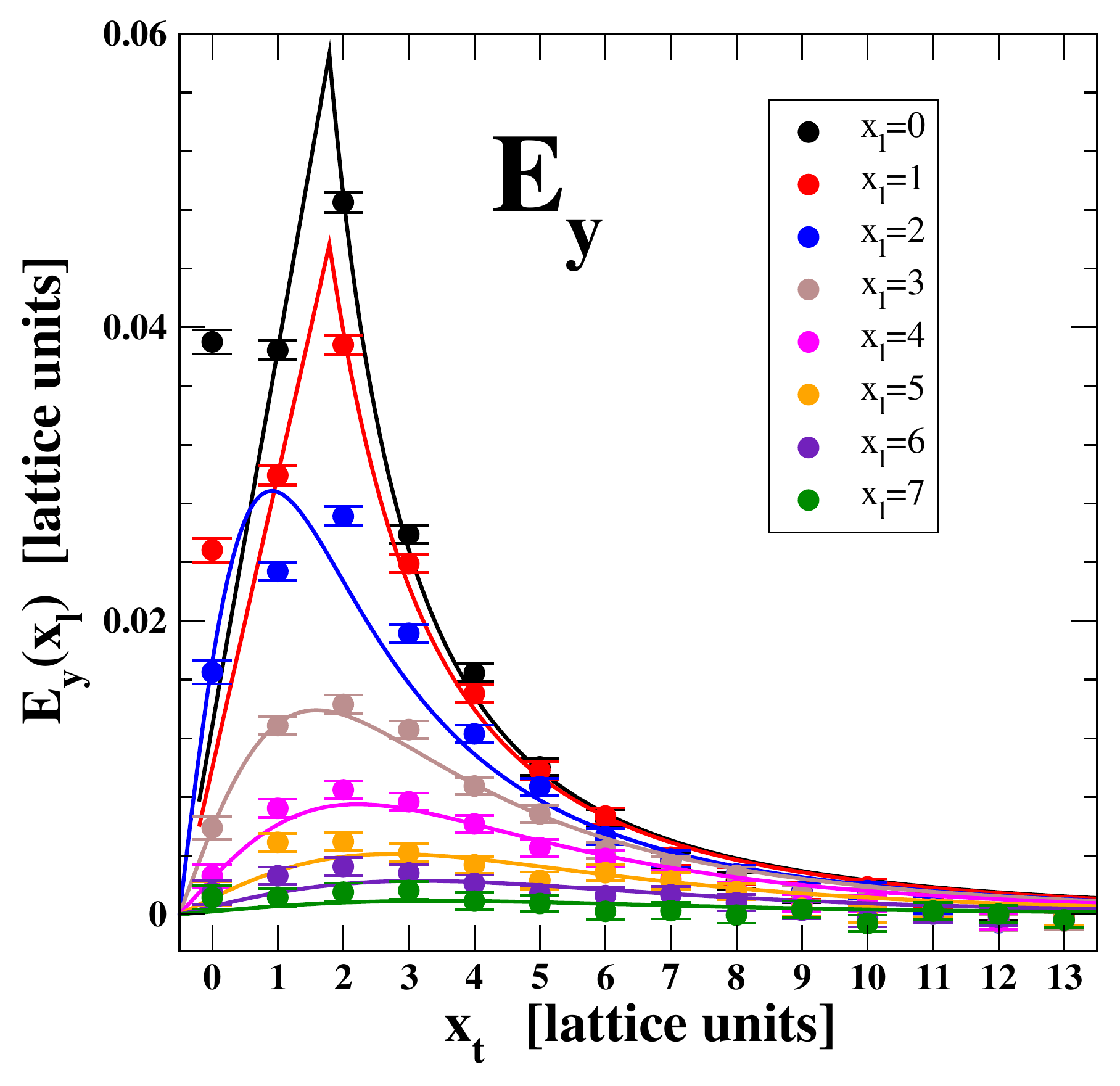}}\vfill
   \subfigure[]%
             {\label{fig:CoulombEz}\includegraphics[width=0.9\columnwidth,clip]{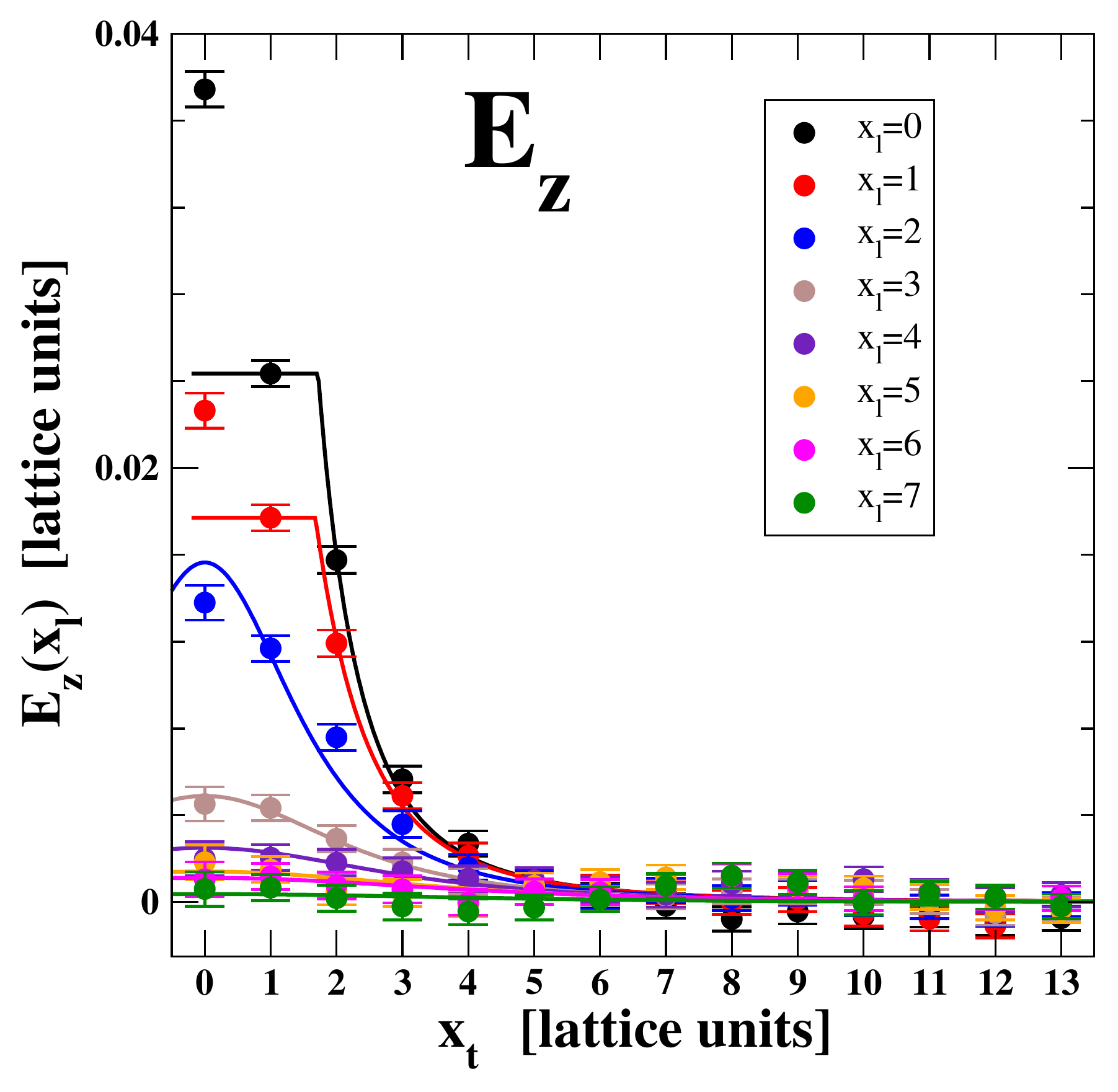}}
\caption{$E_y(x_t)$~\protect\subref{fig:CoulombEy} and $E_z(x_t)$~\protect\subref{fig:CoulombEz} and in lattice units at
  several values of $x_l$ at $\beta=6.240$, $d=16a\simeq1.14$ fm, together
  with the Coulomb fit according to Eq.~(\ref{C2}).}
\label{fig:CoulombEyz}
\end{figure}

  In this Appendix we give some details about the fit of
  the transverse components of the chromoelectric field with the Coulomb
  law given in Eq.~(\ref{C2}). For definiteness, we concentrate on the
  case $\beta=6.240$ and distance between the sources equal to $d=16a
  \simeq 1.14$~fm. The other cases considered in this work were treated
  similarly.

\begin{figure}[htb] 
\centering
\includegraphics[width=0.9\columnwidth,clip]{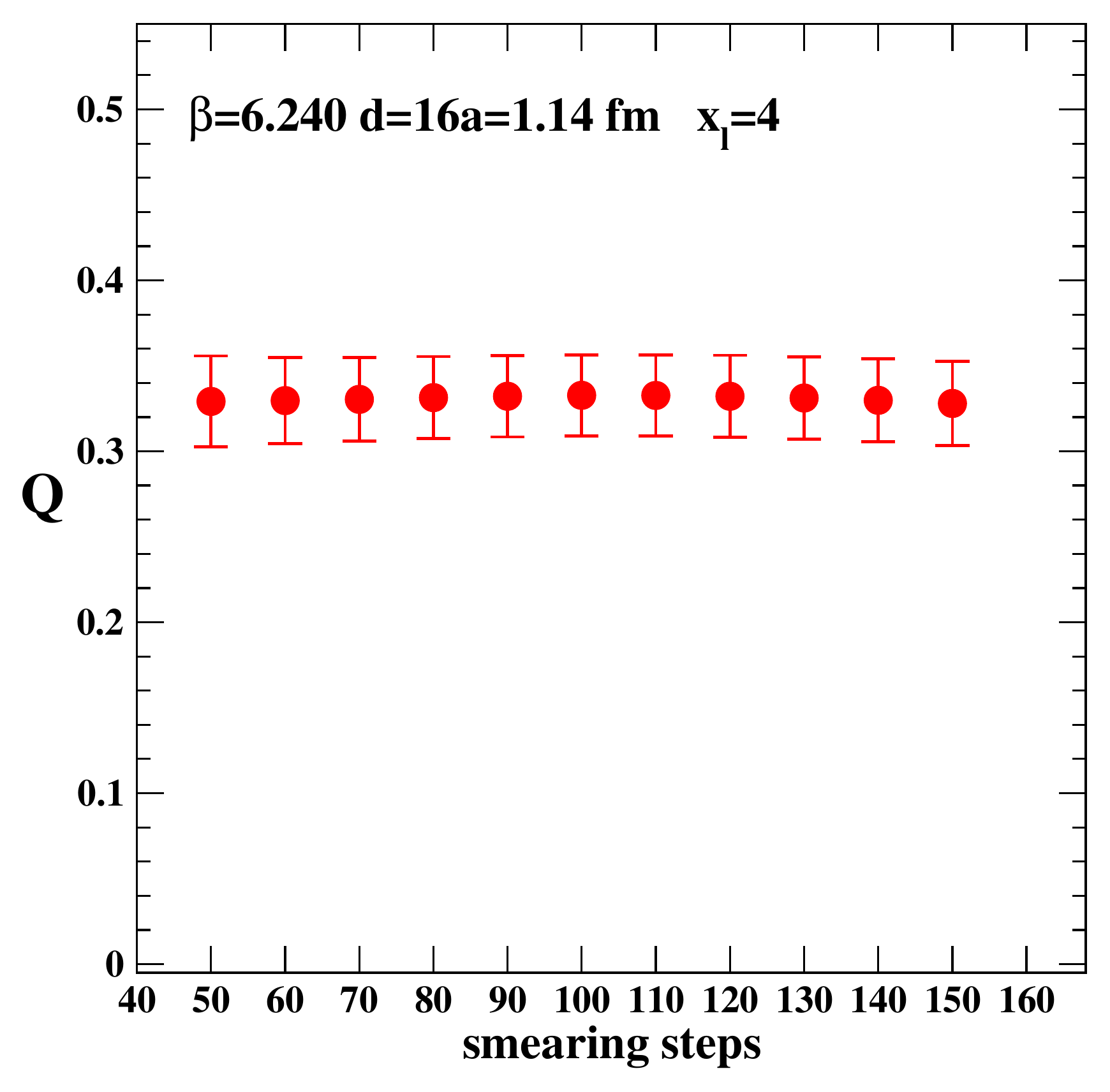}
\caption{$Q$  from the Coulomb fit to $E_y$ field at  $\beta=6.240$, $d=16a\simeq1.14$ fm, and
$x_l=4a$ versus smearing steps.} 
\label{fig:Q_vs_smearing}
\end{figure}
%

%
\begin{table}[htb]
\begin{center}
\caption{Values of the fit parameters extracted from Coulomb fits of the $E_y$ field component
at $\beta=6.240$ and $d=16a\simeq1.14$ fm.
}
\label{tab:coulombFitEy}
\newcolumntype{Y}{>{\centering\arraybackslash}X}
\begin{tabularx}{0.9\linewidth}{@{}|Y|Y|Y|Y|@{}}
\hline\hline
$x_l$ &  $Q$          &  $R_0$ [lattice units] & $\chi^2_r$  \\ \hline
0         &  0.308(5)   &   2.287(24)                &  4.3             \\
1         &  0.312(6)   &   2.498(32)                &  4.0             \\
2         &  0.301(8)   &                                   & 14.3            \\
3         &  0.333(15) &                                   &  2.8             \\
4         &  0.333(24) &                                   &  2.6             \\
5         &  0.313(37) &                                   &  2.3             \\
6         &  0.305(44) &                                   &  1.6             \\
7         &  0.269(86) &                                   &  0.9             \\
\hline\hline
\end{tabularx}
\end{center}
\end{table}
%

%
\begin{table}[htb]
\begin{center}
\caption{Values of the fit parameters extracted from Coulomb fits of the $E_z$ field component
at $\beta=6.240$ and $d=16a\simeq1.14$ fm.
}
\label{tab:coulombFitEz}
\newcolumntype{Y}{>{\centering\arraybackslash}X}
\begin{tabularx}{0.9\linewidth}{@{}|Y|Y|Y|Y|@{}}
\hline\hline
$x_l$ &  $Q$             &  $R_0$ [lattice units]  & $\chi^2_r$  \\ \hline
0         &  0.281(16)   &   1.792(41)                 &  1.6             \\
1         &  0.292(21)   &   2.018(60)                 &  0.9             \\
2         &  0.355(25)   &   2.422(65)                 &  0.6           \\
3         &  0.279(26)   &                                   &  0.2             \\
4         &  0.341(54)   &                                   &  0.6             \\
5         &  0.397(102) &                                   &  1.1             \\
6         &  0.626(191) &                                   &  0.6             \\
7         &  0.488(432) &                                   &  0.9             \\
\hline\hline
\end{tabularx}
\end{center}
\end{table}

  In Fig.~\ref{fig:CoulombEyz} we compare the profiles of the
  $y$- and $z$-compo\-nents of the chromoelectric field on the transverse planes
  labeled by $x_l=0, \dots, 7$ with fitting curves of the form given
  in~(\ref{C2}). The field components were obtained after 100 smearing
  steps.

  The fit parameters extracted from the Coulomb fit to $E_y$
  and $E_z$ on transverse planes at $x_l \leq 7$
  are summarized in Tables~\ref{tab:coulombFitEy} and~\ref{tab:coulombFitEz}.
  For $x_l=0,1$ the point at $x_t=0$ has been excluded from the fit. 
  
  In order to obtain the final values for $Q$ (as well as for $R_0$), as quoted in     Table~\ref{tab:coulombFit},
  only Coulomb fits whose quality is higher than $10\%$, on datasets where the error on the
  largest extracted value of the field is $<20\%$, have been taken into account.
  Then weighted averages have been computed, along with the corresponding statistical error and a systematic uncertainty to account for the variability of $Q$ and $R_0$ among different acceptable fits.

  The behavior of the $Q$ value so determined under smearing
  is shown in Fig.~\ref{fig:Q_vs_smearing}.


\end{document}